\documentstyle[12pt,aasms4]{article}

\lefthead{Ali et al.}
\righthead{Delayed Millimetric Emission from GRBs}

\begin{document}


\title{A SEARCH FOR MILLIMETRIC EMISSION FROM GAMMA RAY BURSTS}

\author{S. Ali, R. K. Schaefer, M. Limon, and L. Piccirillo}

\affil{ Bartol Research Institute, University of Delaware, Newark, DE 19716 }

\begin{abstract} 
We have used the 2- year Differential Microwave Radiometer data from the 
COsmic Background Explorer (COBE) satellite to systematically search for 
millimetric (31 - 90 GHz) emission from the Gamma Ray Bursts (GRBs) in 
the Burst And Transient Source Experiment (BATSE) GRB 3B catalog.  The large 
beamsize of the COBE instrument ($7^\circ$ FWHM) allows for an efficient 
search of the large GRB  positional error boxes, although it also means that 
fluxes from (point source) GRB objects  will be somewhat diluted. A likelihood
analysis has been used to  look for a change in the level of millimetric emission
from the locations of 81 GRB events during the first two years (1990 \& 1991)
of the COBE mission. The likelihood analysis determined that we did not find
any significant millimetric signal before or after the occurance of the GRB. We
find 95\% confidence level upper limits of 175, 192 and 645 Jy or, in terms of
fluxes, of 9.6, 16.3 and 54.8 $\cdot 10^{-13}$ erg/cm$^2\ s$,  respectively at 
31, 53 and 90 GHz.
We also look separately at different classes of GRBs, including
a study of the  top ten (in peak flux) GRBs, the ``short burst" and ``long
burst" subsets,  finding similar upper limits.  While these limits may be
somewhat higher than one would like, we estimate that using this technique with
future planned missions could push these limits down to $\sim$ 1 mJy.  
\end{abstract} 



\section{Introduction}

The first detection of Gamma Ray Bursts (GRBs) were made by the VELA  nuclear
monitoring satellites in the 1960s (\cite{Kle73}). After the launch of the
COMPTON Observatory with the Burst and Transient Source Experiment (BATSE), the
number  of detected GRBs has greatly increased.  Despite the added knowledge, 
we still do not know what the source of GRBs are.  One piece of information
which could help us to unravel the puzzle would be to have observations at
other wavelengths, so we can identify a population of counterparts for separate
study.  Despite many different counterpart searches over different  portions of
the radio spectrum, no GRB counterpart population has been
identified (\cite{Dessenne}, \cite{Frail}, \cite{Koranyi},
\cite{Hjellming}, \cite{Schaefer}, for a review see \cite{McNamara}). 
The previous cited works were concentrated
on studying the delayed emission of single GRB events.
Two works (\cite{Inzani} and \cite{Baird}) have been published so far about
an extended sky search on radio-counterpart similar to the work presented in
this paper.

Two problems arise when searching for counterparts: 1) the error boxes on
GRB locations from BATSE are typically a few degrees, requiring a large amount 
of searching time with usual astronomical telescopes, and 2) the spectrum, 
intensity, and duration of the counterpart signals are all very model 
dependent, so it is hard to optimize counterpart search experiments.  
We note that the extreme isotropy of the GRB population (\cite{bri96}) 
suggests that the GRBs are cosmological sources.  In some cosmological models 
where the GRBs are powered by cosmic fireballs, the GRBs will manifest delayed 
emission at energies much lower than gamma ray.  However the timescale when 
the lower frequency emission peaks can be anywhere from
hours to months after the GRB, depending on the geometry and other properties
of the burst source (e.g., \cite{pacr93,mnr93,mnr97,katz94}).  

Here we focus on a search for delayed emission or precursor from GRBs in the 
millimetric region of the spectrum.  This is a more thorough and complete 
analysis than was presented earlier in (\cite{scha95}).  Millimetric 
observations were made fortuitously by
the COBE (COsmic Background Explorer) satellite with the Differential Microwave
Radiometer (DMR) experiment.  COBE observes the full sky every 6 months at 3 
frequencies (31, 53, and 90 GHz) with 6 radiometer pairs.  (For a more complete 
description of the COBE mission, see e.g., \cite{bog92}.)   For 
models of GRBs with delayed millimetric emission, the best constraints for the 
whole GRB population can be found only from the COBE database.

    Unfortunately for this study, COBE was not designed for studying point
sources.  The radiometer horns have a FWHM of $7^\circ$ to purposely dilute the
effect of point sources.  The radiometers have rms noises of $\sim 23-59$ mK 
$\sqrt{\rm sec}$ in antenna temperature.  1 mK corresponds to a point source 
emission of $\sim$ 1800 $(\nu/53$ GHz$)^2$ Janskys. In 1 second of
observations, COBE has the following sensitivities: 26.4, 28.8 and 98.6 KJy
for, respectively, the 31, 53 and 90 GHz channel. This sensitivities are of the
same order or magnitude of the two similar high sky coverage searches by
\cite{Baird} and \cite{Inzani}. Given these noise levels, the 
best use of COBE data is for studying delayed or anticipated emission on 
long observing
times where the noise can be reduced by integration to a useful level.  The
COBE  data has  also been used to put upper limits on prompt emission which 
would be coincident with the GRB event (see \cite{bone95}).  Another way to 
beat the radiometer noise is to
average over a large ensemble of GRB.  We have done our analysis on the data 
from the first two years of publicly available COBE data.  Only 207 GRBs were 
recorded by GRO during this period.  This number can be increased by a factor 
of $\sim 4$ by using the remaining two years of COBE data.  We plan to analyze
the remaining two years in the future. 

The study presented in this paper produced upper limits which are not very
stringent when compared with some theoretical scenario like, for example,
\cite{pacr93}. The same authors, however, encourage pushing the search for
radio counterpart to higher frequencies. According to their model, the 
peak flux expected is proportional to $\nu^{5/8}$. This paper present a search
for counterparts at the 3 highest frequencies ever used. Other observers rarely
worked in the GHz range. Another important issue concerns the fact that the
GRBs may be strongly beamed. The radio beaming factor $\Gamma_{radio}$ 
increases as $\nu^{9/16}$ and the likelihood of detecting radio transients is
therefore maximized by observing at the highest possible frequencies.

\section{Data Selection}
\subsection {GRB Selection Criteria}

    We have had to decide the amount of sky which is included in our
``observation" of the Gamma Ray Burst location.  We need to balance two
conflicting trends.   The accuracy of our sky measurement, determined by $N$
observations, goes as $N^{-1/2}$, so we want to observe a large area to 
increase accuracy.  On the other hand, If we observe too large a
region of sky we will dilute any signal with observations of blank sky.  
Intuitively, one might expect that the observations will be optimized if we 
collect
observations in an solid angle area of radius $R$ given by $\pi R^2  =  
2\pi( \sigma^2_g + \sigma^2_c)$, where $\sigma_g^2$ is the error 
angle of the GRB location and $\sigma_c= 3^\circ$ is the Gaussian width of the
COBE DMR beams.  We have used the BATSE team's recommended error angle which is
determined by the quadrature sum of the statistical error angle and a
1.6$^\circ$ systematic error angle.  A more formal argument verifying that
choice of radius $R$ 
is indeed optimal is given in the appendix.  

   With the proper radius of coverage around the GRB central position chosen, 
we must then decide what range of gamma ray burst error boxes to
include in our analysis.   The radii of the statistical error circles range
from a low of $0.28$ degrees to a high of $27$ degrees.  There are two
problems with very large error circles. First, averaging over a large areas 
will mean that any GRB signal will be severely diluted, and so will not
contribute much useful information.  Second, larger error circles will include
lots of radio sources (galaxies, AGN, etc.), and so the variance of the 
background subtracted data does not really decrease with increasing error 
circle size as argued above.  We have decided to carry out our analysis only 
for those GRBs with statistical error radii less than 3.5 degrees, the half
width at half maximum 
of the COBE beamsize.  At this angle, the signal is diluted by a factor of 
two when compared to that from a GRB with zero statistical position error.  
Including these bursts with larger error angles leads to no significant 
improvement on our limits on the GRB emission, so we do not consider them 
here any further.  The GRB locations and error angles are taken from the BATSE
3B catalog (\cite{Batse3b}). 
    
\subsection{Selection of COBE Data}

   The COBE team has provided the DMR observations as a set of time ordered
data.  This set contains the temperature differences observed by 
each pair of radiometer horns integrated for each half second of the flight. 
The average positions are given for each horn, and information about the
quality of the data, and also whether any contaminating sources were in the 
field of
view.  We eliminated any data contaminated by transient solar system sources, 
{\it e.g.}, the moon or the planets, but we allow all other data including 
regions of the galaxy because they are presumably reasonably static on time 
scales of days.  We also eliminate any data flagged by the COBE team as 
``bad data" for a variety of reasons, ({\it e.g.}, level spikes, calibration
testing, bad attitude information, {\it etc}.) 

\section{Background Subtraction and Analysis}

  Since we are using the COBE time ordered data set, we must apply corrections
to the data which were needed for the COBE team to construct the DMR skymaps.  
The first and most significant of these is the correction for the magnetic 
susceptibility of
the Dicke switches in the radiometers.  Although Kogut et al (1996) report
that this correction is small ({\it e.g.}, $\stackrel{<}{_\sim} 15$ $\mu$K 
rms) when averaged over the entire 
map, the effect can be quite large in a single 0.5 second observation.  For
example, we found days when the 53A GHZ channel had corrections as large as 
$\sim 0.6$ mK at certain locations along COBE's Earth orbit.  

The next correction we applied was the correction for the local dipole induced
by the satellite motion around the Earth and the Earth around the sun.  This
is particularly important for looking at low declination point sources as 
this will induce a maximal 6 month signal change in the measurements.  This 
happens because the
satellite's observing strategy allows these GRBs to be seen only 
during two different times during the year: when the Earth is moving toward 
the GRB and when the Earth is moving away from the GRB.  If we do not subtract
this effect we see an apparent signal with a 6 month periodicity in the GRB
observations.  

  COBE measures the difference in antenna temperature between the GRB position
and some reference point 60$^\circ$ away.  This angle defines a ring of
comparison points.  COBE orbits along the Earth's terminator, and as the Earth
moves in its orbit, the COBE viewing pattern sweeps through the sky (and hence
also the reference ring).  This reference ring is not uniform; in addition to
containing gradients induced by the dipole anisotropy, it usually
slices through a portion of the galactic disk, which is a strong radio
emitter.  This observing pattern therefore introduces a time dependent
background noise into the GRB observations as COBE sweeps through the different
portions of the reference ring.   

  The static background radio emission can be estimated using the 
synthesized skymaps.  The COBE team has collected the observations and 
constructed six maps of the sky, one for each of the radiometer pairs at 
each of the three frequencies.  Every time a GRB location is observed we 
subtract from the signal the temperature difference from the two horn 
locations, determined by the average temperature of the two maps from the two 
same frequency radiometer pairs.  One must know whether the horn observing 
the GRB is given a weight $+1$ or $-1$ when calculating the difference.  In 
figure \ref{grbtemp}a, we show daily averages of the observed raw 
temperatures, as well as the expected background signal expected from the 
skymap.  When the background is subtracted from the raw temperatures we get 
the residual signals seen in figure \ref{grbtemp}b.  We have checked that the
remaining residuals are indeed consistent with Gaussian random noise, as
expected if there is no signal. 

    We want to use these residuals to look for delayed emission or precursors
from the GRB source.  Unfortunately, we must have some model in mind to test
the  data for goodness of fit.  The model cannot be very complicated for
testing against the limited sample here, because most of the GRB positions are
only observed part of the year, and the time interval between the GRB and the
end of the COBE data is different for each burst.  When the full data set is
analysed, we can test more sophisticated models, but for now we model the
millimetric GRB  signal as a simple increase in emission coincident with the GRB
event and which  remains at a constant level thereafter.   

\section{Results}

   We have found the differences of the average signals coming from the GRB
location before and after the GRB event for all six DMR receivers 
({\it e.g.}, 31A, 31B, 53A, {\it etc.}).  We then make a
$\sigma_{channel}^{-2}$ weighted average of the A and B channels 
for a final result at each frequency.  This is especially important for the 
31 GHz channel as the noise in the 31B GHz channel increased by a factor 
of more than 2 on 14 October 1991.  Since this increase happens in the time 
{\sl after} most of the gamma ray bursts, weighting the averages by 
$\sigma_{channel}^{-2}$ ensures that the excess noise will not lead to false 
detections at 31 GHz.     We also find the weighted 1 $\sigma$ error on
the differences.  We then calculate the likelihood that the ensemble of
differences in temperature contain a significant GRB signal.  

  We have looked at the results of the entire set and three interesting
subsets.  We present the results for each of these in the next four
subsections.

\subsection{All GRBs with error angle radii $\leq$ 3.5$^\circ$}

    We look at the complete set of GRB bursts in the BATSE 3B catalog and
select all of those with statistical error angle radii less than 3.5$^\circ$.  
This
leaves us with 81 GRBs which have temperature differences before and after the 
event.  
The likelihood functions are shown in figure \ref{grblike}a, and the 95 \%
confidence upper limits are shown in Table 1.  These limits cover the ensemble
of GRB events.  We note that including all 81 GRBs leads to a bump in the
likelihood of the 31 GHz frequency which might be construed to be a slight
detection.  After looking at the temperature differences, we find that this
``detection" is due largely to three GRBs (910712, 910718, and 910807 in the
BATSE 3B catalog).  Two of these, 910712 and 910807, have somewhat incomplete
coverage leading to larger variability in both the GRB averages and the 
background level derived from the skymap.  The coverage after the
GRB event is particularly spotty in 910807.  The daily temperature averages 
for these GRBs at 31GHz are shown
in figure \ref{oddballs}.  It can be seen in figure \ref{oddballs}b that GRB 
910718 has more complete coverage than the other two odd events, but we find
that the temperature is larger before the event than after the event. 
Unless this GRB has a higher amount of emission before the GRB event than 
after, there is no GRB signal here.  Certainly no clear emission features are 
evident in these data.  

    If these were real detections, we might expect that they would be seen also
in at least the 53 GHz and possibly the 90 GHz channels.  However, all of these
higher frequency temperature differences for these ``odd" GRBs are consistent 
with zero at the 2 $\sigma$ level, so we do not believe these are real
detections.  We also calculate what the 31 GHz likelihood (in figure
\ref{grblike}a) and 
limit are if we ignore these ``odd" GRBs.  This limit is denoted by ``w/o 3" 
The limits obtained from the 53 and 90 GHz channels
are little changed as these GRBs were already consistent with zero in these 
other channels.  Lastly we note that there is another ``odd" GRB seen at 53
GHz (see the temperature history in figure \ref{oddballs}d).  Although the 53 
GHz likelihood for all GRBs is completely consistent 
with no detection, GRB 910626 shows up as a negative detection particularly in
the top ten and short burst subsets.  If we eliminate this burst from the 53
GHZ likelihood, the upper limit on the whole set at 53 GHz drops to 192 Jy.  

    We also note that if all of the channels had the same detector noise 
temperature, we would expect that the limits would scale as $\nu^2$ when we 
convert from temperature
to flux.  However, since the noise varies by a almost a factor of 3, we see
that the limits from the 53 GHz (quietest channel) observations are nearly the 
same as those from the 31 GHz (noisiest channel) observations.

\subsection{Bursts longer and shorter than 2 Seconds }

     We now divide the GRB population into two subsets based on the GRB 
duration.  It has been shown that there are two seemingly different populations 
(\cite{kouv93}) which are identified when their $T_{90}$, parameter, {\it i.e.} 
the length of time in which 90 \% of the flux is observed, is longer or 
shorter than 2 seconds.  The short bursts have harder spectra. 
Within the context of a cosmological GRB interpretation, it is
expected that the short bursts are nearby and the long bursts are more 
distant (see, e.g., \cite{bri95}).  Separating these classes might 
reveal some feature diluted when the populations are combined.  

    Most of the GRBs with error angles $\leq 3.5^\circ$ are long bursts.  This
is a consequence of the fact that it is easier for BATSE to get better location
information when the burst occurs over a longer time.  There are 72 long 
bursts and 9 short bursts in our data set.  The likelihood functions for the 
long bursts are shown in figure \ref{grblike}b.  Since 
most of the bursts in our selected data are long, it is not surprising that 
the upper limits from the long bursts are very similar to those for the 
entire set.  In the total set we found 3 GRBs which caused a seeming false 
detection and all of these are long bursts.  The limits and likelihood are
calculated excluding those GRBs, just as described above, and shown in figure
\ref{grblike}b and table 1.  

  There are only 9 short bursts with error angles $\leq 3.5^\circ$ in this GRB
sample.   Since these bursts are usually interpreted as being closer, it is 
hoped that better model limits may be obtained.  The likelihood functions for 
this set are shown in figure \ref{grblike}b.  In this case, we see a weak 
detection occurs in the 53 GHz channel.  This bump seems to be entirely due 
to 910626, whose temperature history is shown in figure \ref{oddballs}b.  We 
can see that the observations
are extremely sparse for this burst.  As with the three ``odd" GRBs found in 
the long burst sample, this event does not show a significant temperature
difference in the 31 and 90 GHz data, so we doubt that this is a real
detection.  By throwing out this GRB we improve the 95 \% confidence limit at
53 GHz by about 30 \%.  However, because of the small number of GRBs in this 
sample, throwing away one significantly affects the limits in the other
channels.  In table 1, we show how the limits weaken in the other channels when
we throw away 910626.  We have only shown the difference in the 53 GHz
likelihood function before and after disposing of 910626 in figure 
\ref{grblike}c.

\subsection{ The Top Ten GRB}

   Finally, we select the top ten GRBs in peak flux seen by BATSE during the
first two years of COBE data.  One might hope that since these GRBs had the
strongest gamma ray emission that they might also show the strongest radio
effect.  Since these were very strong bursts, enough photons were observed
to fix accurate positions, so the error radii of these bursts are rather small. 
As with the other sets we show the likelihood functions in figure 
\ref{grblike}d and the
95 \% confidence upper limits are given in table 1.  As in the short burst set
the 53 GHz channel shows a weak detection.  Once again this seems to be due to
the burst 3B 910626 which is common to both subsets.  If we again remove 
3B 910626 from the sample, the 53 GHz
likelihood function drops dramatically and the 53 GHz upper limit decreases by
30\% (see table 1).  However, in this case dropping one of the 10 GRBs from the
sample does not affect the upper limits for the 31 and 90 GHz channels.  The 90
GHz limit increases by an amount consistent with $\sqrt{10/9}$, but the 31 GHz
limit is nearly unchanged.  

\section{Discussion and Future Prospects}

   As we find no significant detections of GRBs, the upper limits present in 
Table 1 are our main results.  We can look in detail at the results from the
subsets to see if they are consistent with observing simply noise.  The main
difference between the subsets is the number of GRBs used.  The
scaling with $1/\sqrt{n}$ is seen in the data, however there is certainly more
going on there.  This is most noticeable when comparing the limits derived in
the short burst and top 10 subsets as the limits are much larger in the
short burst set.   
 
   Part of the reason for this can be discovered when we look at the average
GRB error angles in the different subsets given in table 1.  The average error
angle for the whole set is $1.7^\circ$; the angles for the short bursts are
significantly larger than the average and the angles for the top 10 bursts are
significantly smaller than average.  This means there will be different amounts
of point source dilution of the signal in the different subsets.  Using the
proper beam dilution factors when comparing the results accounts for most of
the rest of the difference.  For example, the difference between the limits 
from the 53 GHZ channel between the short and top ten subsets is explained very
well by the difference in GRB number and dilution factors.  This is 
reassuring, as
53 GHz is the quietest channel.  Interestingly, the limits are more consistent
if we consider only those in which the ``odd" GRBs are excluded.  Lastly, the 
90 GHz channel limits for the short bursts are somewhat larger than the scaling
relations would predict.  

 Limits on GRB emission from the total population are roughly what we would 
expect if the  detectors were seeing nothing but detector noise.  We take 
this as an indication that method of ``observing" GRBs is working quite well.  
Since GRO was only launched in April 1991, the overlap between COBE and GRO is 
only 8 months during the first two years of COBE data.  Therefore, 
including the final two years of COBE data will quadruple the number of GRBs
observed, so the sensitivity would be increased by a factor of 2.

  A much larger increase in sensitivity can be achieved if currently planned
satellites INTEGRAL (a gamma ray observatory satellite) and MAP and/or
COBRAS/SAMBA (microwave anisotropy observing satellites) are successfully 
deployed.  We estimate that combining observations from INTEGRAL with either 
MAP or COBRAS/SAMBA will increase the sensitivity in the 20-90 GHz region by 
3 orders of magnitude or more.  The improvements come largely from three 
factors: 1) The detector noise has decreased by about two orders of magnitude, 
2) INTEGRAL will provide error boxes on the order of arcminutes so less blank
sky will need to be observed, and 3)
the smaller beamsizes add another order of magnitude in sensitivity 
to point sources.  (Note that although the solid angle is over two orders of 
magnitude smaller than COBE's, one order of magnitude is lost because the 
satellites will spend less time observing the GRB location.)   
   We point out that COBRAS/SAMBA also has a cryogenic bolometer detector at
higher frequencies which should yield considerably better results.  The best 
limit should be obtained from the 143 GHz channel of COBRAS, which has the 
lowest detector noise.  We estimate that if there is no signal from GRBs, this 
channel will yield an upper limit of about 1 mJy.

\section{Conclusions}

We have analyzed the set of GRBs observed by BATSE on GRO for delayed
emission or precursors in the 31-90 GHz region using the data from COBE.  We
find  limits for the amount of possible millimetric emission from this set using
a crude model (D.C. change in emission level) of the millimetric emission.  Our
main results are the likelihood functions in figure 3 and the 95\% confidence
upper limits on GRB emission in Table 1.  For the entire set of 81 GRBs with
useful positional error boxes (error radii $\leq 3.5^\circ$) the 95\%
confidence upper limits are 175, 192, and 645 Jy at 31, 53, and 90 GHz,
respectively.  

   We have also selected several subsets for study.  We have looked at 
the long and short bursts ($>$ or $<$ 2
seconds in duration) subsets, as it has been speculated that they are at 
different distances in cosmological models.  We also examine the top
ten GRBs (rated by BATSE peak flux counts) under the reasoning that if there is 
any effect it is likely that the effect will be greatest in the strongest 
GRBs.  We find that these subsets do not show any significant detections,
and that the limits scale properly from the entire population limits.  

  These limits are the best on the delayed/precursors flux in the millimetric
region of the spectrum.   While the particular limits from this study perhaps
do not put much pressure on theoretical models, we have demonstrated the
viability of the using this technique.  We expect that using this method with
the future INTEGRAL in combination with 
MAP and/or COBRAS/SAMBA satellites will improve the limits by many orders of
magnitude, which will be much more interesting theoretically.  Lastly we
emphasize that with these satellite, we have complete sky coverage, so the
entire population of GRBs can be observed.  
Seeing the entire population allows comparisons of
different GRB subsets, which is useful for diagnosing the properties
of the GRBs.  We believe this technique therefore has a promising future. 

 \acknowledgements

 We want to thank A. Kogut, D. Leisawitz, and J. 
Newmark for help in deciphering the COBE data and related software.  We also 
thank Lyman Page for supplying us with estimates of detector noise and 
beamsizes for the MAP satellite.  We gratefully acknowledge NASA for its 
support in an ADP grant NRA-93-OSS-95. 

\appendix 

\section{Sky Averaging Dilution Effects}

     We show here that there is an optimal angular distance over which one
should integrate to get the best limit on GRB emission.   We first want to
consider the signal seen by a point source GRB at location ${\bf r_g}$, when
the COBE beam center is pointing at a position ${\bf s}$.  The temperature
$T({\bf s})$ seen by the COBE DMR radiometer is then
\begin{equation}
T({\bf s}) = C {\rm exp}\left[ - {|{\bf s} - {\bf r_g}|^2 \over 2 \sigma_c^2}
\right],
\end{equation}
where $C$ is the flux of the GRB source converted into antenna temperature, and
$\sigma_c = 3^\circ$ is the COBE Gaussian beamsize.  

   The GRB location ${\bf r_G}$ is not known accurately; instead we are give a
Gaussian probability of finding it at a location ${\bf r}$ within an error 
angle of $\sigma_g$ of the central location, 
\begin{equation}
{\rm Pr_G}({\bf r}) = {1 \over 2 \pi \sigma_g^2} {\rm exp}\left[ - {|{\bf r} 
- {\bf r_g}|^2 \over 2 \sigma_g^2} \right],
\end{equation}
where $\sigma_g$ is the quadrature sum of the statistical
error ($\sigma_{stat}$) and the systematic error ($\sigma_{sys}=1.6^\circ$), as
recommended by the BATSE team \cite{Batse3b}. 
If we define the origin so that ${\bf r}_g=0$, the expected temperature seen by
pointing the radiometer at location ${\bf s}$ is
\begin{eqnarray}
\langle T({\bf s}) \rangle &=& C \int {d^2r \over 2 \pi \sigma_g^2} \ 
{\rm exp}\left[ - {|{\bf r}|^2 \over 2 \sigma_g^2} \right]
{\rm exp}\left[ - {|{\bf s} - {\bf r}|^2 \over 2 \sigma_c^2} \right],
\nonumber \\
&=& {C\over \sigma_g^2/\sigma_c^2 + 1}\ {\rm exp}\left[ - {|{\bf s}|^2 \over 
2 (\sigma_c^2 + \sigma_g^2)} \right].
\end{eqnarray}

  In order to get a measurement we average the observations made in a circle
or radius $R$ around the GRB central location.  The sampling of the sky in this
circle is roughly uniform, so the result of our analysis yields an unweighted
average of the expected temperature $ \bar{\langle T \rangle} $ in this circle:
\begin{eqnarray}
\bar{\langle T \rangle} &=& {C\over \sigma_g^2/\sigma_c^2 + 1} 
\int {d^2s \over  \pi R^2}\  {\rm exp}\left[ - {|{\bf s}|^2 \over 
2 (\sigma_c^2 + \sigma_g^2)} \right],  \nonumber \\
&=& {C\over \sigma_g^2/\sigma_c^2 + 1} {1\over x^2} \left[1 - {\rm exp}(-x^2)
\right],
\label{observtemp}
\end{eqnarray}
where $x^2 = R^2/[2 (\sigma_g^2 + \sigma_c^2)]$.  We now can calculate the 
the amount of signal dilution to expect when we make our average of GRB
``observations".   If we were pointed directly at a GRB location which had no
position error, we would get an antenna temperature of $C$. The ratio of the
term on the RHS of equation \ref{observtemp} to $C$ gives us a dilution factor
$D$:
\begin{equation}
D = {1\over \sigma_g^2/\sigma_c^2 + 1} {1\over x^2} \left[1 - {\rm exp}(-x^2)
\right],
\label{dilution}
\end{equation}
which describes the diminishment of any GRB signal in our averaging process.  
The resulting difference of averaged antenna temperatures before and after the
GRB event must be divided by this factor for each GRB.  

    We can also use equation \ref{observtemp} to define an optimal angle over
which to average our temperature observations.  As we increase the radius of
our observing circle $R$, the instrumental noise in the 
average of GRB observations goes down like the inverse square root of the 
number of observations $N^{-1/2}$ or like $R^{-1}$.  On the other hand 
increasing $R$ also implies that we will be diluting the signal.  The dilution
factor in equation \ref{dilution} is roughly constant when 
$R^2 \stackrel{<}{_\sim} \sigma_c^2 +
\sigma_g^2$ and decreases as $R^{-2}$ at larger $R$.  The signal to noise ratio
of our measurement will then be determined by the ratio of the dilution 
factor to the noise.  At small $R$, $S/N \propto R$ and at large
$R$, the $S/N \propto R^{-1}$.  The maximum ratio occurs then
when $R = 1.02\ [ 2\ (\sigma_c^2 + \sigma_g^2)]^{1/2}$.  This is the angular
size of the circles we used for making our GRB averages.  

   Here we also can see explicitly the dependence of the dilution factor on 
the error angle of the GRB.  If we double the maximum size of the error angle 
we allow for our GRB to be included, say $7^\circ$ instead of $3.5^\circ$, for 
the optimal value of $x$, the signal will be diminished by a factor of 
$\sim 1/4$.   Increasing the area searched also allows for more noise sources 
to be included in the average, and this can be seen to a small degree in the 
Long Burst/Short Burst subset comparison.

\clearpage

\section*{FIGURE CAPTIONS}

\figcaption {\label{grbtemp} 24 hour average Antenna temperatures measured by 
the COBE satellite.  Here we show the
observations by averaging both 53 GHz receiver pairs at the location of the 
GRB 3B 910609, for which we have good coverage over the whole two years.  
Panel a) shows the raw antenna temperatures and the solid line is the 
background signal predicted using the average of the 53 A and B skymaps.  
Panel b) shows the residual antenna temperatures, which are still largely 
dominated by Gaussian
receiver noise.  Large deviations from zero in the figure occur when only a
few observations were made in that 24 hr period and so are very noisy.  }

\figcaption { \label{grblike} The Likelihood functions for all of the GRB and 
for various subsets.  Panels a), b), c), and d) are for the total, long, 
short, and top ten burst sets, respectively.   The think lines are the 
likelihood functions including all GRB in the sample.  The thin lines are 
used to show the changes when we drop certain GRBs from the sample (see text).}

\figcaption {\label{oddballs} Shown are the 24 hour averages vs. time of the 
GRBs which are responsible for the weak ``detections" in our likelihood 
functions.  The panels a) b) c) and d) are the temperature histories
for the labelled bursts. } 

\clearpage

\begin{table*}
\begin{center}
\begin{tabular}{ccccc}   \tableline
	Channel & All & Long  & Short & Top 10 \\ \tableline
	
	31 GHz & 247 & 331 & 562 & 382 \\
	
	53 GHz & 217 & 202 & 990 & 755 \\
	
	90 Ghz & 645 & 675 & 2205 & 1172 \\ \tableline
              &        &        &           &          \\      
   $<$ error angle $>$\tablenotemark{a} & 1.7$^\circ$ & 1.6$^\circ$ 
& 2.2$^\circ$    & 0.71 $^\circ$ \\
              &        &        &           &          \\      
 \tableline
 \multicolumn{5}{c}{ Improved Limits Removing Selected Bursts (see text)} \\
\tableline
                     & w/o 3\tablenotemark{b}  &  w/o 3 & &\\
	31 GHz & 175 &  185  & &\\ \tableline
               & w/o 910626     &         & w/o 910626   &  w/o 910626 \\
	31 GHz & & & 712 &  375 \\  
	53 GHz & 192 & & 690 &  555 \\  
	90 GHz & & & 2700 &  1225 \\ \tableline

\end{tabular}
\end{center}

\label{tbl-1}


\caption{ Upper limits (95\% confidence level) on delayed or anticipated 
millimetric emission (in Jy) for each of the various subsets.}

\tablenotetext{a}{Statistical error circle radius averaged over the subset}
\tablenotetext{b}{Limit with out including the 3 GRBs which imply a detection 
at 31 GHz (see text).}

\end{table*}


\begin{thebibliography}{}

\bibitem[Baird et al., 1975]{Baird} Baird, G. A. {\it et al.}, 1975, Ap.
J., 196, L11

\bibitem[Boggess et al. 1992]{bog92} Boggess, N. {\it et al.}, 1992, ApJ, 397, 
420

\bibitem[Bontekoe et al 1995]{bone95} Bontekoe, T., Winkler, C., Stacy, J.G., 
\& Jackson, P.D., 1995, Astrop. \& Sp. Sci., 231, 285

\bibitem[Briggs et al 1996]{bri96} Briggs, M., {\it et al.}, 1996, ApJ, 459, 40

\bibitem[Briggs et al 1995]{bri95} Briggs, M., 1995, Astrop. \& Sp. Sci., 231, 3

\bibitem[Dessenne et al., 1996]{Dessenne} Dessenne, C. A. {\it et al.},
1996, MNRAS, 281, 977

\bibitem[Frail et al., 1994]{Frail} Frail, D. A. {\it et al.}, 1994, Ap.
J., 437, L43

\bibitem[Hjellming \& Ewald, 1981]{Hjellming} Hjellming, R. M. \& Ewald S. P.,
1981, Ap. J., 246, L137

\bibitem[Inzani et al., 1982]{Inzani} Inzani, P. {\it et al.}, 1982, AIP
Conf. Proc. Vol. 77, p. 79

\bibitem[Katz 1994]{katz94} Katz, J. L., 1994, ApJ, 432, L107

\bibitem[Klebesadel Strong \& Olson 1982]{Kle73} Klebesadel, R. W., Strong, 
I.B., \& Olson, R. A., 1973, ApJ, 182, L85

\bibitem[Kogut et al 1996]{COBE96} Kogut, A., {\it et al.}, 1996, 
COBE Preprint 96-10.

\bibitem[Koranyi et al., 1995]{Koranyi} Koranyi, D. M. {\it et al.}, 1995,
MNRAS, 276, L13

\bibitem[Kouveliotou et al 1993]{kouv93} Kouveliotou, C., {\it et al.}, 1993, 
ApJ, 413, L101 

\bibitem[McNamara et al., 1995]{McNamara} McNamara, B. J. et al., 1995,
Ap. J., 452, L25

\bibitem[Meegan et al 1996]{Batse3b} Meegan, C., et al., 1996, unpublished. 
The 3B catalog and preliminary analysis can be found on the World Wide Web at 
URL http://heasarc.gsfc.nasa.gov/cossc/cossc.html

\bibitem[Meszaros \& Rees 1993]{mnr93} Meszaros, P. \& Rees, M.~J., 1993, 
ApJ, 418, L59

\bibitem[Meszaros \& Rees 1997]{mnr97} Meszaros, P. \& Rees, M.~J., 1997,
ApJ, 476, 232

\bibitem[Paczynski \& Rhoads 1993]{pacr93} Paczynski, B. \& Rhoads, J.~E., 
1993, ApJ, 418, L5

\bibitem[Schaefer et al 1995]{scha95} Schaefer, R.K., Ali, S., Limon, M., 
\& Piccirillo, L., 1995,  Astrop. Sp. Sci. 231, 331

\bibitem[Schaefer et al., 1989]{Schaefer} Schaefer, B. E. {\it et al.},
1989, Ap. J., 340, 455

\end{thebibliography}
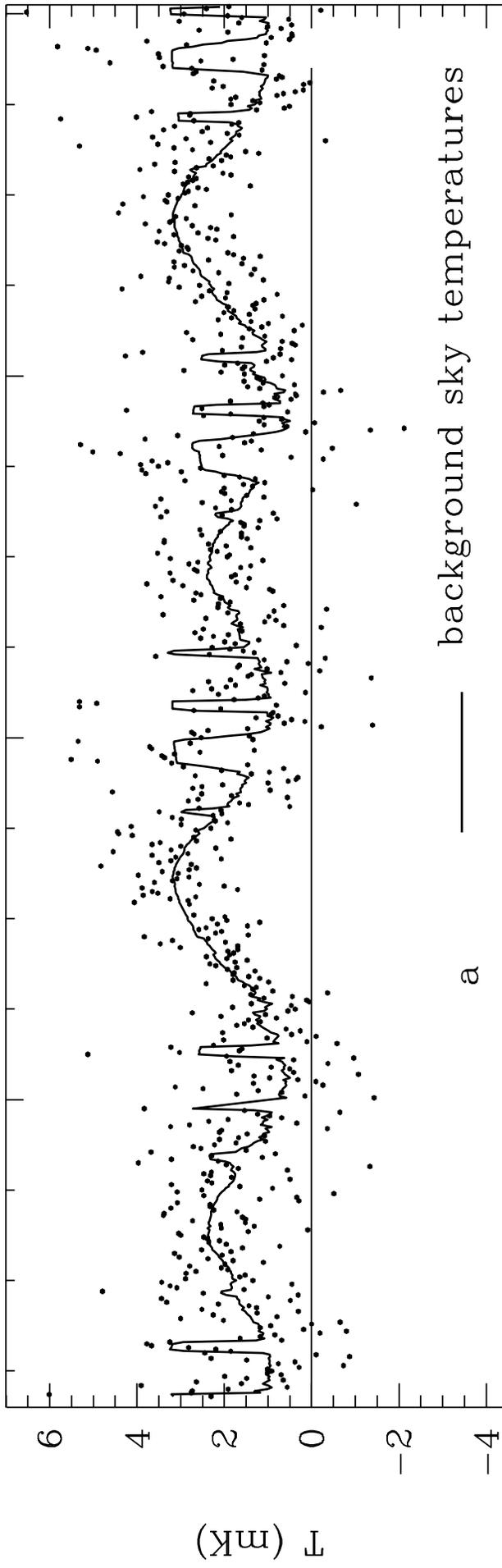
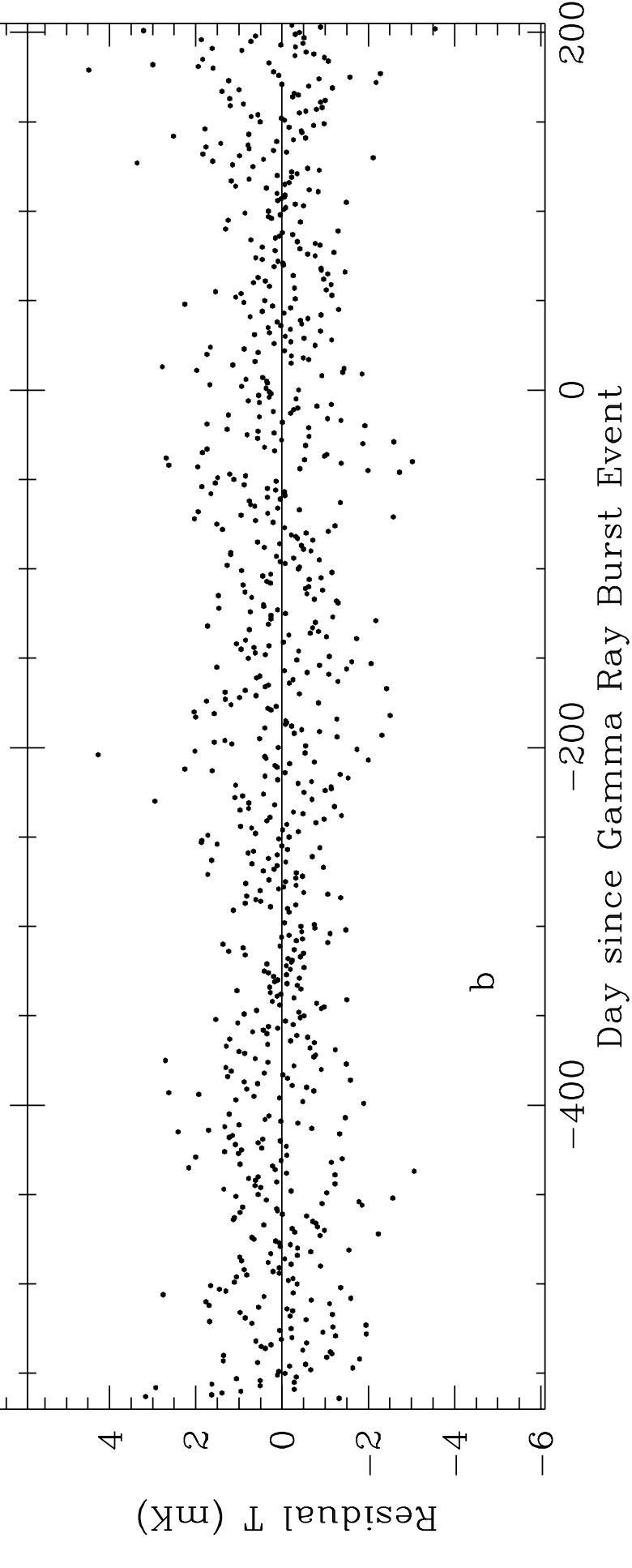
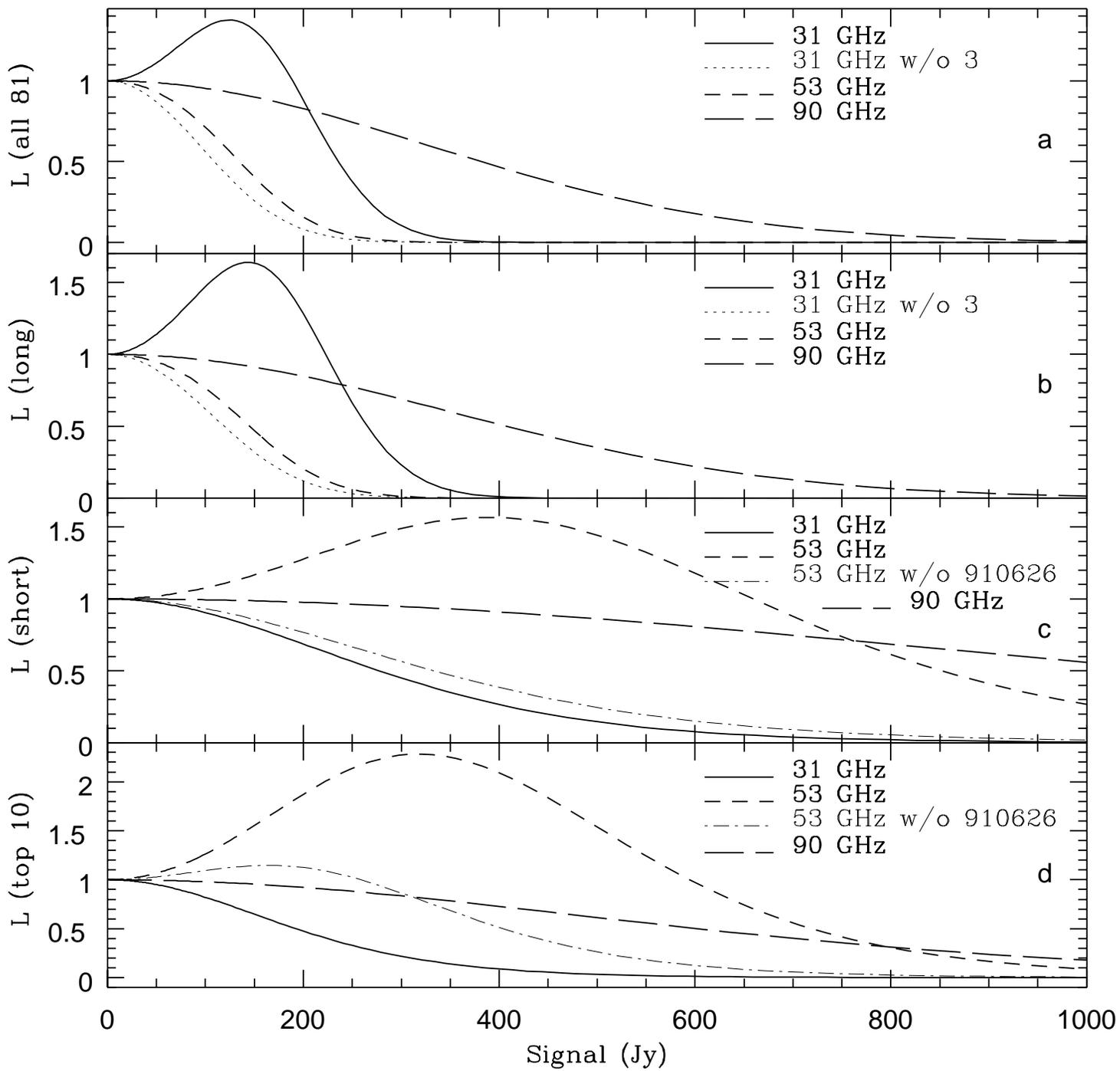
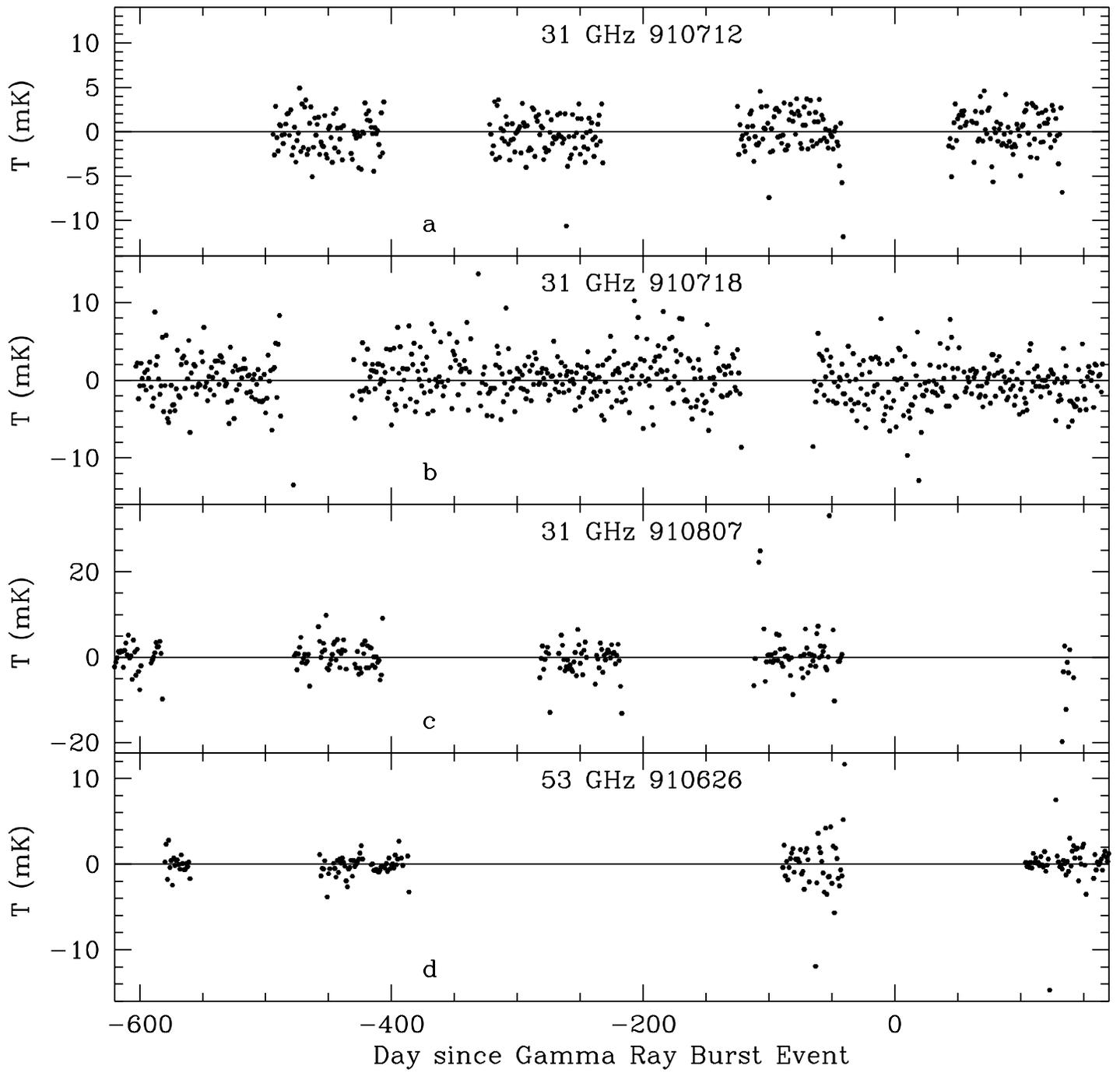
\end{document}